\documentclass[prl,aps,superscriptaddress,twocolumn]{revtex4-1}
\usepackage{siunitx}
\usepackage{graphicx} 
\graphicspath{{graph/}}
\usepackage[usenames]{color}
\usepackage{amsmath,amssymb}
\usepackage{gensymb}
\usepackage{natbib}
\usepackage{xcolor}
\def\strutdepth{\dp\strutbox}
\def\nw#1{\strut\vadjust{\kern-\strutdepth\vtop to0pt{\vss\hbox to\hsize
{\hskip\hsize\hskip5pt$\leftarrow$\hss\strut}}}{\em #1}}

\begin{document}

%Title of paper
\title{Dynamic solid surface tension causes droplet pinning and depinning}

\author{M. van Gorcum}
\affiliation{Physics of Fluids Group, Faculty of Science and Technology, Mesa+ Institute, University of Twente, 7500 AE Enschede, The Netherlands.}
\author{B. Andreotti}
\affiliation{Laboratoire de Physique Statistique, UMR 8550 ENS-CNRS, Univ. Paris-Diderot, 24 rue Lhomond, 75005, Paris.}
\author{J. H. Snoeijer}
\affiliation{Physics of Fluids Group, Faculty of Science and Technology, Mesa+ Institute, University of Twente, 7500 AE Enschede, The Netherlands.}
\author{S. Karpitschka}
\affiliation{Max Planck Institute for Dynamics and Self-Organization} %Maybe Stefan can put the full text here.

\date{\today}

\begin{abstract}
The contact line of a liquid drop on a solid exerts a nanometrically sharp surface traction. This provides an unprecedented tool to study highly localised and dynamic surface deformations of soft polymer networks. One of the outstanding problems in this context is the stick-slip instability, observed above a critical velocity, during which the contact line periodically depins from its own wetting ridge. Time-resolved measurements of the solid deformation are challenging, and the mechanism of dynamical depinning has remained elusive. Here we present direct visualisations of the dynamic wetting ridge formed by water spreading on a PDMS gel. Unexpectedly, it is found that the opening angle of the wetting ridge increases with speed, which cannot be attributed to bulk rheology, but points to a dynamical increase of the solid's surface tensions. From this we derive the criterion for depinning that is confirmed experimentally. Our findings reveal a deep connection between stick-slip processes and newly identified dynamical surface effects.
\end{abstract}
\pacs{}

\maketitle

Liquid drops on vertical glass windows frequently get stuck due to sub-micrometric heterogeneities of the surface. Indeed, while a drop on a perfectly flat homogeneous surface will slide down under its own weight, surface defects result into pinning of the contact line and can give rise to a characteristic stick-slip motion. Different aspects of the depinning transition on rigid surfaces have been studied experimentally and theoretically~\cite{RevBonn09}, revealing critical phenomena~\cite{Rosso02} that are blurred by thermal activation at the nanoscale \cite{PLDRA16}. From an engineering perspective, there is a continued effort in designing surfaces with low contact angle hysteresis for purposes of hydrophobicity, self-cleaning, or anti-fouling coatings~\cite{solomon2016lubricant,quere2008wetting}.

Recently, wetting of soft surfaces has generated a large interest: The liquid-like surface properties of reticulated polymer networks and brushes can offer nearly hysteresis-free substrates \cite{Lhermerout2016aa,snoeijer2018paradox}. However, despite the absence of hysteresis, drops on these soft surfaces \emph{do} exhibit a stick-slip motion when forced to spread beyond a threshold velocity~\cite{Kajiya2013a,Kajiya2014aa,C7SM01408B,KarpNcom15}. This stick-slip behaviour is highly unexpected, since the steadily moving contact line is not pinned to a material point of the solid, but ``surfs" a wetting ridge created by the capillary forces located at the contact line. This suggests a rather intriguing scenario of self-pinning, and subsequent depinning of the contact line from its own wetting ridge. While stick-slip motion was qualitatively correlated to the viscoelastic rheology of the substrate \cite{Kajiya2013a,Kajiya2014aa}, or weakening of the wetting ridge~\cite{C7SM01408B}, the current models predict a stable steady-state motion even at large velocities~\cite{KarpNcom15}. As such, there is no explanation why the contact line would depin from its own ridge and start a stick-slip cycle.

In this Letter we identify the origin of the stick-slip transition using a direct high-speed visualisation of the full wetting ridge during self-pinning and depinning. The experiments reveal an increase of the solid angle at the ridge tip with contact line velocity, which is interpreted as a dynamical increase of solid surface tensions. We identify the criterion for the instability and as such reveal the origin of the stick-slip motion: the increased solid angle enables the contact line to depin from its ridge and rapidly slide down its own wetting ridge.
\begin{figure}
\includegraphics[width=86mm]{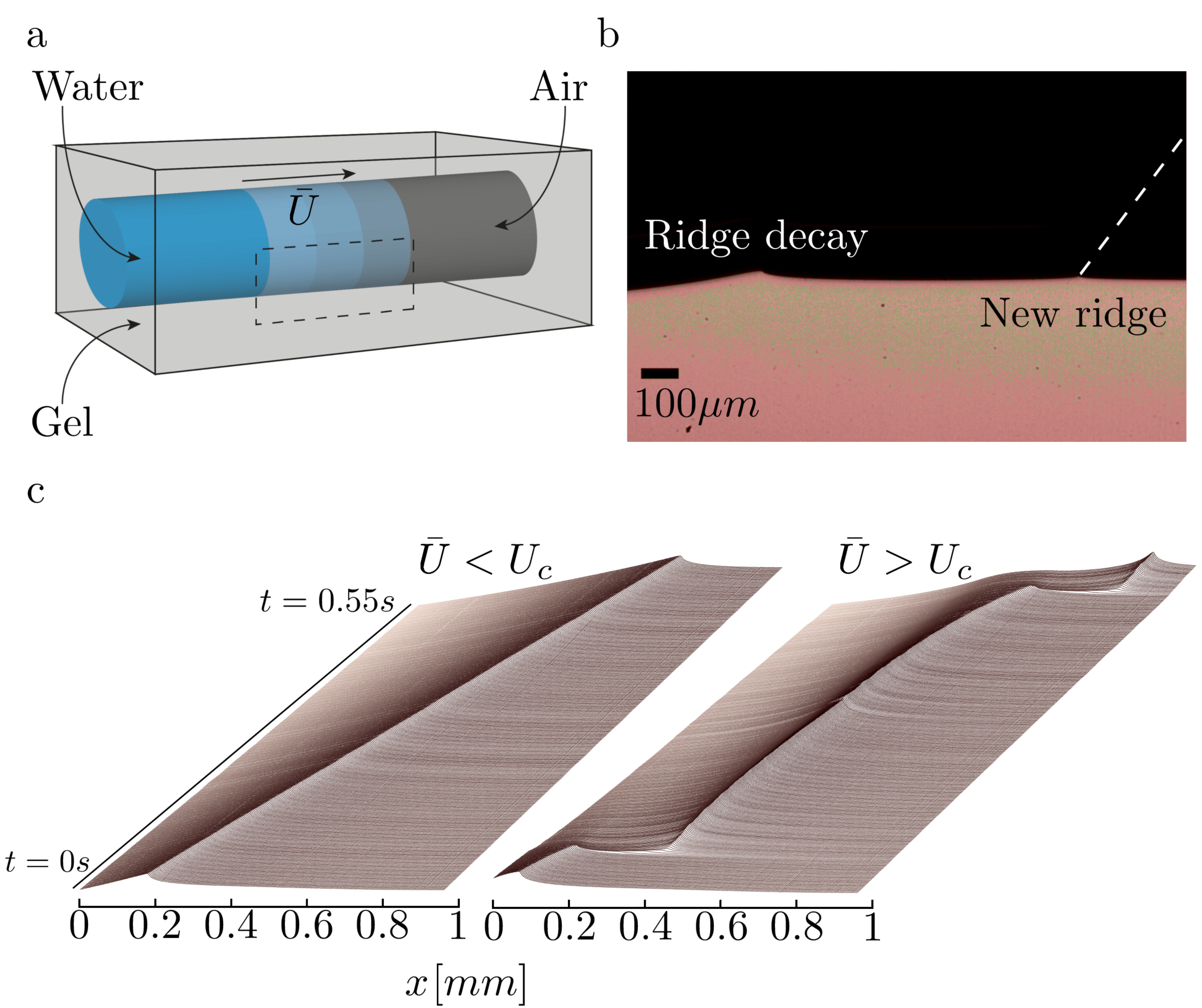}
\vspace{- 4 mm}
\caption{(a) Schematic of the experimental setup. A cylindrical cavity of gel is partially filled with water, with an imposed velocity $\bar U$. (b) The shape of the gel can be accessed optically from the side. The image was taken just after the contact line depinned from the old ridge (left), forming a new ridge (right). The liquid-vapor interface is indicated by the dashed line. (c) Space-time diagram showing the dynamics of the wetting ridge in steady state (left) and in the stick-slip regime, which is reached above a critical velocity $U_c$ (right).}
\vspace{- 5 mm}
\label{Fig1}
\end{figure}

\paragraph{Wetting ridge visualisation~--~}It is notoriously difficult to image the wetting ridge below the contact line. Techniques such as confocal~\cite{xu2017direct} or X-ray microscopy~\cite{PARKNATURE,C7SM01408B} have provided excellent spatial resolution, but the rapid dynamics during stick-slip motion constitute a challenge. To overcome this, we present the setup shown in Fig.~\ref{Fig1}(a) consisting of a square block of transparent gel with a cylindrical cavity. The block has a width of 10 mm, the cavity diameter is 4 mm, leaving about 3 mm of gel thickness. These dimensions ensure a separation of length-scales where the elastocapillary length is small compared to the thickness and the meniscus, while the effect of gravity is still negligible. The cavity is partially filled with MilliQ water. The contact line pulls the gel inwards, creating an axisymmetric wetting ridge. This setup allows to accurately trace the edge of the gel when observing from the side [dashed rectangle in Fig.~\ref{Fig1}(a)] without the otherwise unavoidable optical distortions. Using a high speed camera (3200 frames per second) and a long working distance microscope (giving a spatial resolution of $1.0-2.2\mu$m/pixel) we image  the wetting ridge, and the stick-slip motion with the necessary spatio-temporal resolution. A sample image is shown in Fig.~\ref{Fig1}(b), which was taken during a rapid ``slip" event after the contact line depinned from the ridge. The image reveals both  the new wetting ridge (right) and the decaying old wetting ridge (left). 

The reticulated polymer used is a PDMS gel (Dow Corning CY52-276, mixed at a 1.3:1 (A:B) ratio). The dynamic modulus is accurately described by $G'(\omega)+iG''(\omega)=G[1+(i\omega \tau)^n]$, where the static shear modulus $G=265$~Pa, $\tau=0.48$~s and the exponent $n=0.61$ (cf. Supplementary Information \cite{supplementary}.) With the surface tension of MilliQ water $\gamma=72$~mN/m, this gives rise to the elastocapillary length $\gamma/G = 0.27$~mm, which makes the wetting ridge easily accessible for optical microscopy. We verified that the experimental aspects such as the surface tension of the water and the wetting ridge dynamics remain consistent over extended periods of time (days), even when leaving the water in contact with the PDMS. The motion of the contact line is imposed by filling the cavity at a constant volumetric rate using a syringe pump; control parameter is the imposed averaged contact line speed $\bar U$. The opening angle of the solid ridge was measured at the micron scale within $\pm 3 ^{\circ}$ by fitting the gel profile on both sides of the ridge (cf. Supplementary Information \cite{supplementary}). The contact angle of the liquid was measured in a separate experiment using a sessile drop on a flat surface of the same gel of sufficiently large thickness.

\paragraph{From steady motion to stick-slip cycles~--~}At small velocity $\bar{U}$, the contact line moves in a steady state, as can be seen by the left space-time diagram in Fig.~\ref{Fig1}(c). This steady regime has been studied in previous works \cite{KarpNcom15, Shanahan1995aa, Carre1996a, LALLang96, KarpPNAS16,zhao2018geometrical}, which have shown that the motion is governed primarily by the viscoelastic dissipation inside the solid. The speed at which a drop can spread thus depends on the rheology of the solid, as well as on the size of the wetting ridge where most of the dissipation occurs. 

\begin{figure}
\includegraphics[width=86mm]{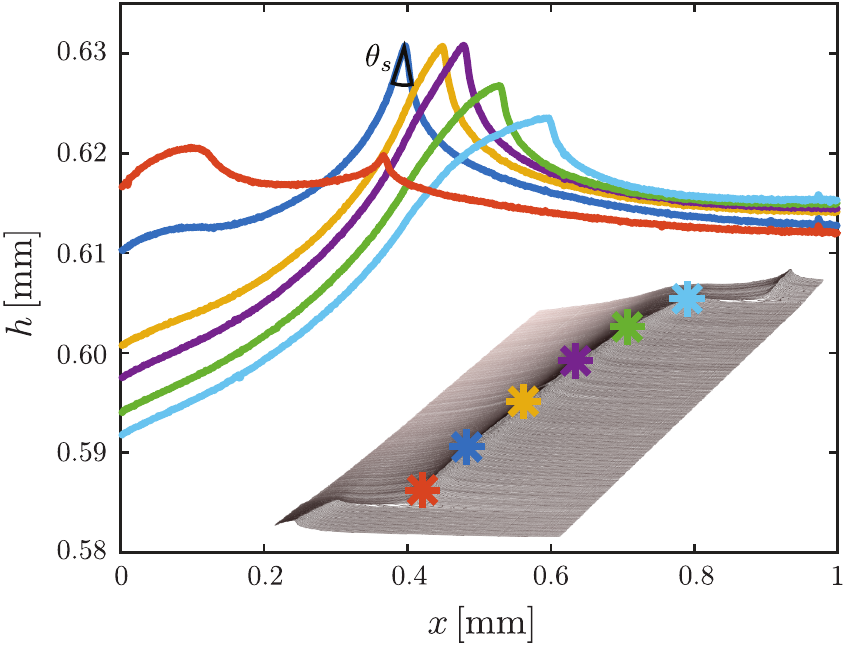}
\vspace{- 4 mm}
\caption{Dynamical ridge shapes at different times during a stick-slip cycle. The coloured asterisks on the inset space-time diagram denote the point in time the snapshot was taken (time interval between snapshots $\approx0.04$~s).}
\vspace{- 5 mm}
\label{Fig2}
\end{figure}

Our prime interest lies in the high velocity regime, where wetting is forced beyond a critical speed $U_c$. The contact line motion then turns unstable, and results into a stick-slip behaviour that we resolve in detail at the scale of the wetting ridge. The space-time plot of a stick-slip cycle is shown in Fig.~\ref{Fig1}(c), with the corresponding dynamic wetting ridges presented in detail in Fig.~\ref{Fig2}. 

At the start of a cycle, the contact line depins from the wetting ridge and suddenly moves with a much higher velocity over the substrate (slip phase). The first data in Fig.~\ref{Fig2} (red) shows the profile just after depinning, where we observe both the abandoned wetting ridge (rounded tip, left) and the newly formed wetting ridge (sharp tip, right). The old ridge is no longer pulled by the contact line and will decay over time. The evolution of the new ridge can be followed by the subsequent profiles in Fig.~\ref{Fig2} (from dark blue to light blue). The wetting ridge initially grows, while also moving along the surface. In contrast to the sudden depinning, the transition to the stick phase is characterized by a continuous deceleration of the ridge. The growing ridge causes more viscoelastic dissipation, slowing down its motion. At a later stage, the wetting ridge becomes markedly asymmetric, with a large rotation of the ridge tip (light blue). Finally, the contact line depins and the stick-slip cycle is repeated. 

The speed during the slip phase is two to three orders of magnitude faster than during the stick phase (Fig.~\ref{Fig1}c). Slip velocities, between $10^2$ and $10^3$~mm/s, are actually comparable to those measured during the wetting of a rigid surface~\cite{RevBonn09,RevSnoB13}. In this phase of rapid motion there is essentially no wetting ridge, and hence negligible viscoelastic dissipation inside the solid. During the slow phase, the velocity typically remains larger than $\approx30\%$ of the critical speed $U_c = 1.0 \pm 0.1$~mm/s. Therefore, the contact line never really ``sticks" to a material point of the substrate, but rather creeps, as for the stick-slip associated with solid friction~\cite{baumberger2006solid}.

The depinning from the wetting ridge cannot be explained as a simple consequence of the viscoelastic ``friction" force. Namely, classical stick-slip in solid friction is due to the sharp decrease of the friction force with velocity, modeled in Coulomb's law by a jump from static to dynamic friction. The decrease of friction causes an overshoot of the speed of the slipping object, which in turn induces the irregular stick-slip regime. However, the gel used in these experiments presents a loss modulus increasing monotonically with frequency, $G'' \sim \omega^n$ with exponent $n=0.61$, so that the dissipation in each material element, and subsequently the effective friction force, increases with velocity. Indeed, theory based on linear viscoelasticity assuming constant surface tensions, predicts a stable, steady solution for \emph{all} velocities \cite{KarpNcom15}. This calls for a more detailed investigation, given below, of how the contact line detaches from the wetting ridge.  
\begin{figure}
\includegraphics[width=86mm]{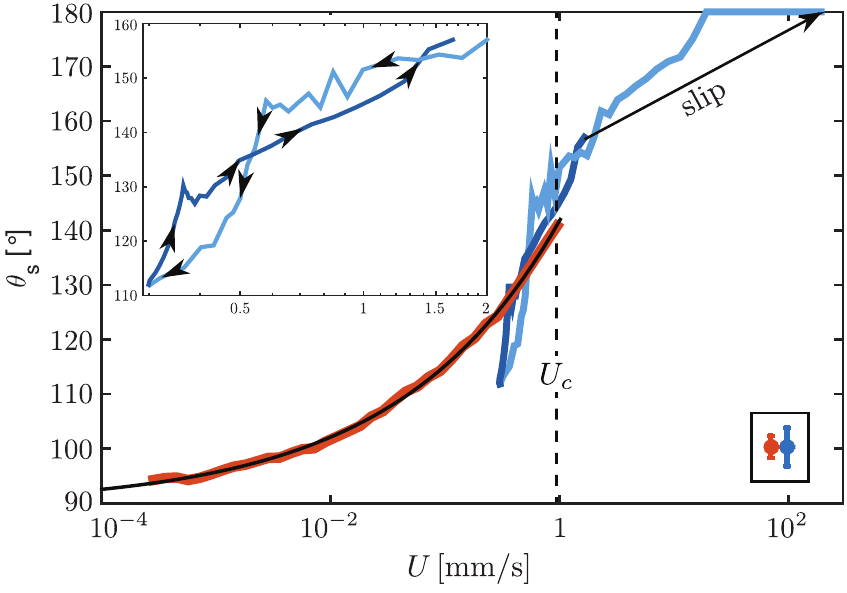}
\vspace{- 4 mm}
\caption{The solid opening angle $\theta_S$ versus contact line speed $U$. The red curve shows $\theta_S$ for the steady state regime ($U<U_c$), accurately fitted by $\theta_S - \theta_{S,0} \sim U^{0.31}$ (black line). The blue curve represents a stick-slip cycle ($U>U_c$). The light blue part of the curve represents the decelerating part of the cycle, where the wetting ridge is growing and the solid opening angle $\theta_s$ is decreasing. The dark blue part of the curve shows the accelerating part of the cycle, up to the point of ridge detachment (slip, arrow). Inset: zoom on the stick-slip cycle with the arrows denoting the direction of time.}
\vspace{- 5 mm}
\label{Fig3}
\end{figure}

\paragraph{Dynamic surface tension~--~}
The main surprise of the profiles in Fig.~\ref{Fig2} is that the solid angle $\theta_S$ at the tip of the wetting ridge  is not constant, but increases with velocity; it will provide the key to the depinning mechanism. Figure~\ref{Fig3} shows the dynamics of $\theta_S$ for both the regime of constant velocity ($\bar{U}<U_c$) and during the stick-slip cycle ($\bar{U}>U_c$). In the steady-state regime (red) we observe a gradual increase with velocity, from the static value $85\pm2\degree$ at equilibrium to $140\pm4\degree$ at the critical speed $U=U_c$. The data is very accurately described by the fit $\theta_S - \theta_{S,0} \sim U^{0.31}$ (black). A similar trend is observed during the stick-slip cycle (blue), where $U$ denotes the instantaneous contact line velocity. The moment of slip is indicated by the arrow, where the surface is very flat and we set $\theta_S$ to $180\degree$. Then, as the contact line velocity decreases, a decrease in $\theta_S$ is observed that follows the same trend as in the steady-state, in first approximation. At the minimum velocity, $\theta_S$ slightly undershoots the steady-state curve in red. Subsequently the contact line velocity increases, while $\theta_S$ follows the steady state curve until depinning occurs and the cycle is repeated.

This unexpected increase of $\theta_S$ with velocity can be interpreted as a dynamical increase of the solid surface tensions. Here we follow~\cite{xu2017direct,xu2018}, where $\theta_S$ is used to measure surface tensions based on the Neumann condition for the contact angles. This is a balance of forces applied to an elementary material system surrounding the contact line: the liquid surface tension $\gamma$ is counteracted by the solid surface tensions, respectively $\Upsilon_{SV}$ (solid-vapor) and $\Upsilon_{SL}$ (solid-liquid). By this argument, we assume that the bulk viscoelastic stress cannot contribute on small scales, which is in fact consistent with the shear-thinning nature ($n<1$) of the gel. This point can be quantified by estimating the bulk stress at a distance $\ell$ from the contact line. The characteristic frequency at which the material is excited by the moving contact line is $\omega\sim v/\ell$, which for the present rheology implies a bulk stress $\sigma \sim G (\tau \omega)^n \sim G (v \tau/\ell)^n$. When computing the integral of viscoelastic stress from a distance $\ell$ up to the contact line, we find $\ell \sigma \sim G (v \tau)^n \ell^{1-n}$; for shear-thinning materials with $n<1$, this integrated stress vanishes in the limit $\ell \rightarrow 0$, so ultimately capillary effects prevail. The cross-over scale from bulk-to-surface is obtained from the balance $\ell \sigma \sim \gamma$, which gives $ \ell \sim (\gamma/G)^{1/(1-n)} (v\tau)^{-n/(1-n)} $.  For the gel used in our experiment, this gives $\ell \gtrsim 10\mu m$ for $U\lesssim 5$ mm/s, ensuring that the measurement of $\theta_S$ is indeed not obscured by bulk effects. Importantly the corresponding characteristic frequencies are within the range that is accessible by bulk rheometry (see supplement~\cite{supplementary}).

\begin{figure}
\includegraphics[width=86mm]{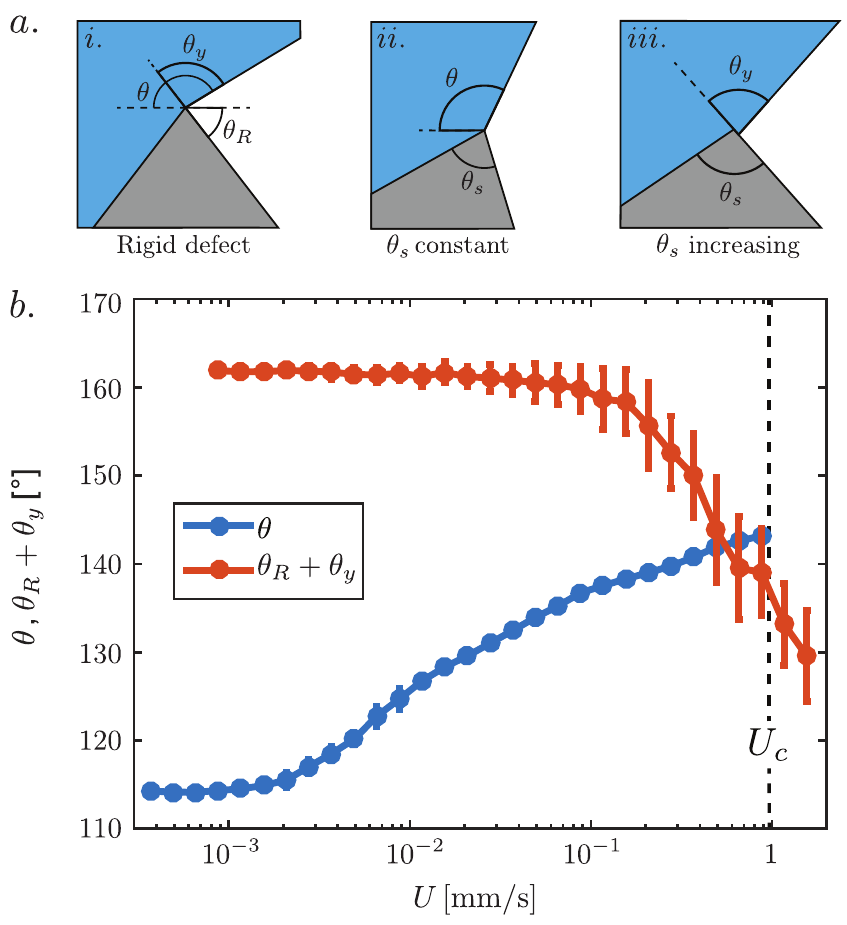}
\vspace{- 4 mm}
\caption{(a). Mechanism for ridge detachment in the stick-slip regime. (i) A rigid defect. The contact line depins as soon as the liquid angle with respect to the inclined edge ($\theta-\theta_R$) exceeds the Young angle ($\theta_Y$). (ii) On a soft wetting ridge with constant surface tensions, the ridge rotates along with any change of the liquid angle ($\theta$), in order to maintain a constant $\theta_S$. (iii) Depinning can only occur when $\theta_S$ increases as a function of $U$. (b). Validation of the depinning condition (\ref{eq:detachcrit}), verified by plotting $\theta$ and $\theta_R + \theta_Y$ versus contact line speed $U$. The critical speed for depinning occurs when $\theta > \theta_R + \theta_y$, at which the contact line can detach from the ridge.}
\vspace{- 5 mm}
\label{Fig4}
\end{figure}

\paragraph{Depinning mechanism~--~}
Finally, we demonstrate how the change in $\theta_S$ is indeed responsible for depinning [Fig.~\ref{Fig4}(a)]. First we revisit the classical depinning from a ``rigid wedge" ~\cite{gibbs1961scientific,oliver1977resistance}, used as a model for topographic roughness on stiff solids [panel (i)]. The contact line can slide off the ridge once the liquid angle \emph{with respect to the inclined edge} exceeds the Young's angle $\theta_Y$ that marks the wetting equilibrium. Any (virtual) displacement to the right with a lower angle will result in a restoring force that pushes the contact line back to the tip of the ridge. 
The depinning criterion thus reads
\begin{equation}
	\label{eq:detachcrit}
	\theta  >  \theta_R + \theta_Y,
\end{equation}
where $\theta$ is the liquid angle with respect to the horizontal, while $\theta_R$ is inclination angle on the right of the wedge [panel (i)]. A similar argument applies for depinning to the left, which in the case of a rigid wedge gives rise a range of possible values for $\theta$. This is the source of contact angle hysteresis on rigid topography, for which (\ref{eq:detachcrit}) goes by the name of the Gibbs inequality \cite{Dyson1988}. For \emph{soft} wetting ridges, however, the mechanics is fundamentally different -- for example, no contact angle hysteresis is observed~\cite{snoeijer2018paradox}. Instead, any change of the liquid angle $\theta$ is followed by a rotation of the ridge itself, as is clearly visible in Fig.~\ref{Fig2}. In the absence of changes of surface tensions, there is no depinning, but instead foresees a stable steady motion at all speeds~\cite{KarpNcom15}. In this case of constant surface tensions, the solid angle $\theta_S$ takes on a constant value (dictated by the Neumann balance), and $\theta$ and $\theta_R$ will always rotate by exactly the same amount to maintain a constant $\theta_S$ [panel (ii)]. By consequence the depinning criterion (\ref{eq:detachcrit}) will never be satisfied. All this changes dramatically, however, when surface tensions allow for an increase of $\theta_S$, and thereby reduce $\theta_R$, enabling the contact line to slide off the wetting ridge [panel (iii)].

This scenario for dynamics-induced depinning is confirmed in Fig.~\ref{Fig4}(b), where we verify the depinning criterion (\ref{eq:detachcrit}). The blue data shows the increase of the liquid angle $\theta$ with velocity $U$. This is compared to the angle relative to the right-side of the ridge, $\theta_R + \theta_Y$, shown as the red data. In accordance with recent work \cite{xu2017direct,schulman2018surface,snoeijer2018paradox}, we set $\theta_Y$ to a constant value measured from equilibrium. The motion leads to a decrease of $\theta_R$ that allows for a crossing of the curves that indeed coincides with $U_c$ [Fig.~\ref{Fig4}(b)]. This is direct evidence for the depinning criterion (\ref{eq:detachcrit}) as the cause of stick-slip. This mechanism does not involve any sudden failure of the Neumann condition, as was suggested in~\cite{C7SM01408B}, which is corroborated by the gradual evolution of $\theta_S$.

\paragraph{Discussion and outlook~--~} 

In summary, we revealed that the surface tension of a soft solid is a truly dynamical quantity, which has important consequences in contact mechanics. It provides the mechanism responsible for depinning and the rapid stick-slip motion of drops on soft substrates. The change in ridge angle $\theta_S$ provides direct evidence for a dynamic coupling between the surface tensions and the mechanical state of the substrate. A similar variability of $\theta_S$, and thus of the solid's surface tensions, was recently observed for \emph{static} drops, when progressively stretching the substrate~\cite{xu2017direct,xu2018}. This was attributed to the so-called Shuttleworth effect~\cite{Shuttleworth1950a,AS16,style2017elastocapillarity}, where the surface tension depends on the elastic surface strain. It is tempting to interpret the results of Fig.~\ref{Fig3} along the same lines, by considering the strain induced by the droplet motion. Indeed, the motion of the contact line induces a rotation (visible in Fig.~\ref{Fig2}) of the wetting ridge with respect to its more symmetric static shape; this rotation leads to an increase of the stretch on the liquid side and a decrease of the stretch on the vapour side. However, the observed change of $\theta_S$ is found to change as $U^{0.31}$, while the rotation angle $\varphi$ -- and its induced strain -- exhibits a much weaker dependence $\sim U^n$ dictated by the rheology exponent $n=0.58$~\cite{KarpNcom15}. It is thus unlikely that the strain-dependence of surface tension is sufficient to explain our observations in Fig.~\ref{Fig3}. 

The above considerations  suggest that surface tension of such a polymer gel is a truly dynamic quantity that depends on the \emph{rate} of strain at the surface as the contact line passes by. This opens the exciting perspective of surface rheological effects, as is for example known for interfaces with surfactants~\cite{fuller2012complex,ritacco2011dynamic}. For the present case of cross-linked polymer networks, this will require a detailed understanding of non-equilibrium interfacial effects that find their origin at a scale smaller than the distance between cross-linkers. The observed time-dependent surface tension suggests, as for thixotropy, the possibility that the surface tension actually depends on a micro-state variable that remains yet to be identified. 
Future work on surface effect in reticulated polymer network will have to focus on the micro-physics of the problem. 

We acknowledge discussions with J. Eggers and A. Pandey. This work was financially supported by the ANR grant Smart and ERC (the European Research Council) Consolidator Grant No. 616918.

\bibliographystyle{apsrev4-1}
%\bibliography{stick-slip}
%

\end{document}

\title{Supplemental Information:\\Dynamic solid surface tension causes droplet pinning and depinning}

\author{M. van Gorcum}
\affiliation{Physics of Fluids Group, Faculty of Science and Technology, Mesa+ Institute, University of Twente, 7500 AE Enschede, The Netherlands.}
\author{B. Andreotti}
\affiliation{Laboratoire de Physique Statistique, UMR 8550 ENS-CNRS, Univ. Paris-Diderot, 24 rue Lhomond, 75005, Paris.}
\author{J. H. Snoeijer}
\affiliation{Physics of Fluids Group, Faculty of Science and Technology, Mesa+ Institute, University of Twente, 7500 AE Enschede, The Netherlands.}
\author{S. Karpitschka}
\affiliation{Max Planck Institute for Dynamics and Self-Organization} %Maybe Stefan can put the full text here.

\date{\today}

\begin{abstract}
In this Supplemental Information we provide experimental details on the measurement of the contact angles of the wetting ridge and on the rheology of the gel. 
\end{abstract}

\pacs{}

\maketitle
\paragraph{Wetting ridge angle measurement~--~}

As a first step, we detect the edge of the gel from images such as shown in Fig. 1 of the main text, using a subpixel measurement scheme. For this, we fit an error function of the grey values of a vertical cut around the edge where the inflection point is set to the subpixel height of the gel. This leads to profiles shown in Fig. 2 of the main text. The imaging scale of our setup is $1.0\ \mu$m/pixel or $2.2\ \mu$m/pixel, depending on the series of experiments, and we estimate the accuracy of the vertical position to be $\approx 0.5\ \mu$m, which is not limited by the imaging scale, but by the optical resolution power and the recording noise.

\begin{figure}
\includegraphics[width=180mm]{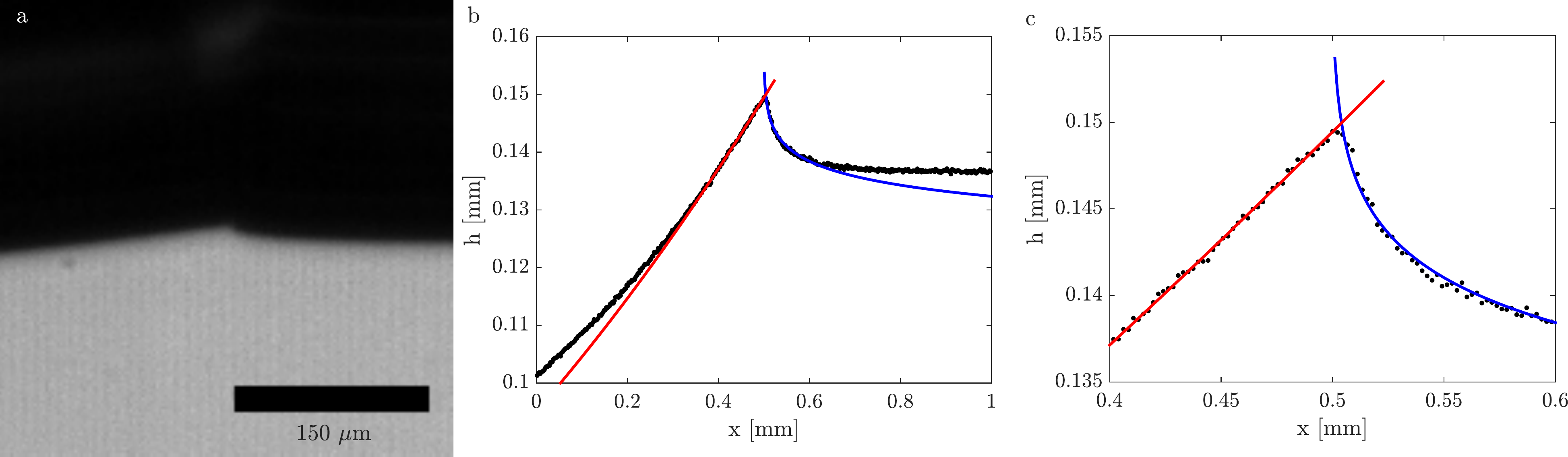}
\vspace{- 4 mm}
\caption{a: Unprocessed image from a measurement series b: Fitting method used to measure the relevant angles on the shown image. On the right (air) side of the gel a generic log function is fit, while on the left side a 2nd degree polyfit is used, c: Zoomed in version of the same data and fit.}
\vspace{- 5 mm}
\label{gelfit}
\end{figure}
Unavoidably, however, the tip of the ridge will appear rounded because of the limited optical resolution of such kind of shadowgraphy setup. The typical residual radius of curvature at the ridge tip that is extracted from the images is about $2-3\ \mu$m, independent of the imaging scale. This indicates that the resolution is limited by the aperture (numerical aperture $\sim 0.1$) of the shadowgraphy setup and not the imaging scale. This optical limitation blurs the discontinuity of the gel's orientation when crossing the contact line. To accurately determine the solid angle $\theta_S$ as reported in Fig. 3 of the main text, we therefore fit both sides of the contact line and subsequently determine the discontinuity from the crossing of the fits (Fig. \ref{gelfit}). To perform the fit, we first estimate the horizontal position of the contact line from the local maximum around the ridge. We then use phenomenological fit-functions that provide an accurate description of the local profile on either side of the contact line. From the profiles in Fig. 2 in the main text, it is clear that the wetting ridges present a strong asymmetry. We therefore used different fits for the two sides. To the right of the contact line (on the side of the air), the shape is fitted by the form $y=a+b \log(c+x)$. To the left of the contact line (on the side of the liquid), a second order polynomial is used, $y=a+b x+c x^2$. For both fits, the points around the maximum within the limit of optical resolution are discarded to account for the tip rounding (1 pixel on each side for the measurement with imaging scale $2.2\ \mu$m). Fig.~\ref{gelfit} of this Supplementary Information shows a representation of the fitting procedure on a measurement with an imaging scale of $2.2\ \mu$m/pixel at $U=0.45$mm/s and $\theta_s = 135^{\circ}$, using 60 datapoints to the left and the right of the detected ridge tip. The fits have a norm of residuals of $5.9 \times 10^{-6}$ for the blue fit and $1.5 \times 10^{-3}$ for the red fit, which are typical values. Panel~c shows a zoom of the of the same fit. 

The intersection of the two fitted curves gives a good estimate of the actual location of the ridge tip, and the slopes at that point then give the relevant angles. Importantly, as described in the main manuscript, the length scale on which the surface curvature could change due to bulk rheological effects is typically more than a factor 10 larger than the optical resolution limit.

\paragraph{Rheology measurement~--~}

\begin{figure}
\includegraphics[width=86mm]{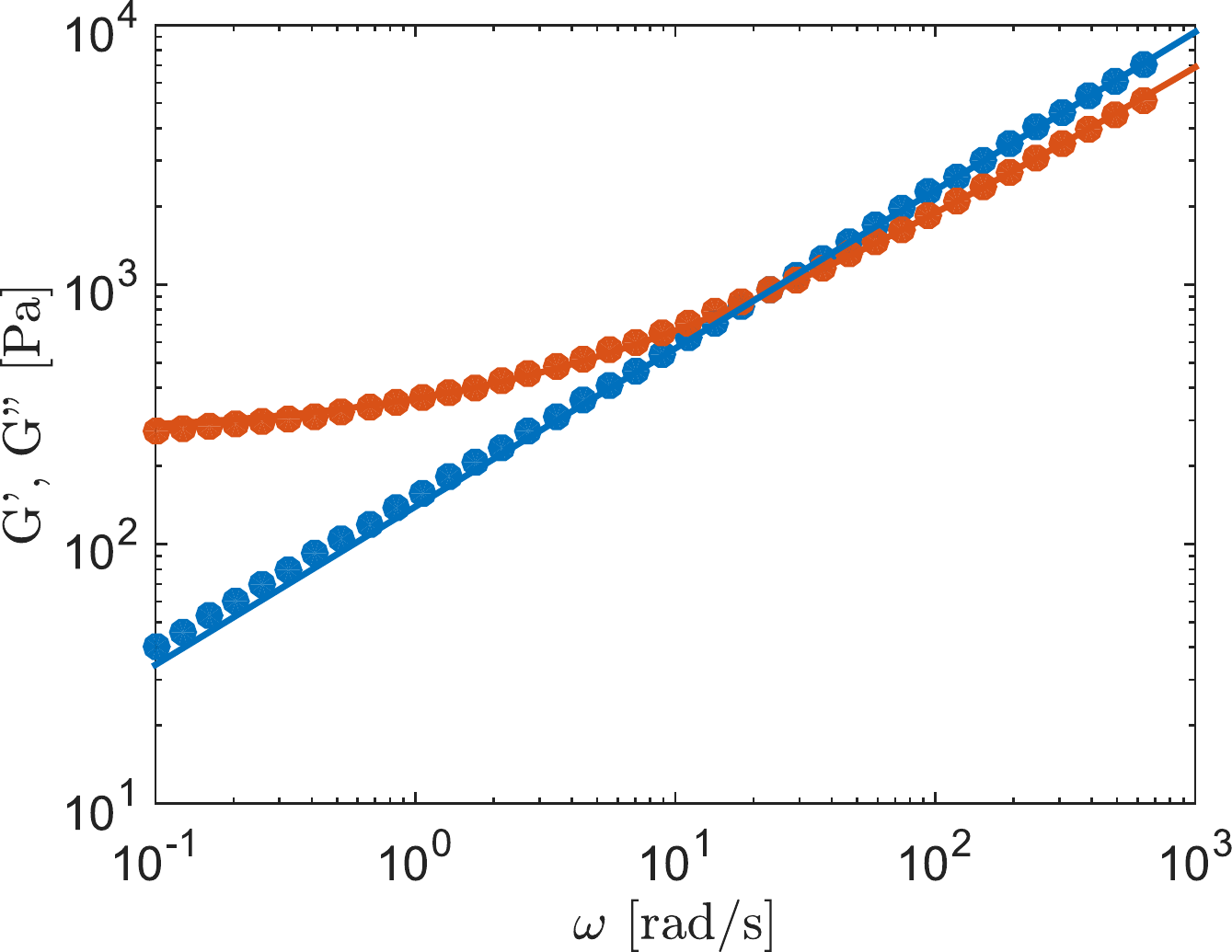}
\vspace{- 4 mm}
\caption{Rheological measurement of the gel, with the fit $\mu(\omega)=G'(\omega) + iG''(\omega) = G(1+(i\omega\tau)^n)$. The best least-squares fit is obtained with $G=265$ Pa, $\tau=0.48$ s, $n$=0.61.}
\vspace{- 5 mm}
\label{rheology}
\end{figure}

The rheological properties were measured with an Anton Paar MCR 502 rheometer in a parallel plate geometry using a frequency sweep at 1\% strain, the results are shown in Fig.~\ref{rheology} of this Supplement. The datapoints are fitted with $\mu(\omega)=G'(\omega) + iG''(\omega) = G(1+(i\omega\tau)^n)$, where $G'$ is the storage modulus and $G''$ is the loss modulus. Using a least-squares fit on the three parameters, we found the static shear modulus $G=265$ Pa, the timescale $\tau=0.48$ s, and the exponent $n=0.61$, as reported in the main text. The fits are superimposed in Fig.~\ref{rheology}.

